\documentclass{aa}

\begin{document}

\title{Dyadosphere bending of light}
\author{V. A. De Lorenci
\inst{1}, N. Figueiredo \inst{1},
H. H. Fliche \inst{2} and
M. Novello \inst{3}}

\institute{Instituto de Ci\^encias --
 Escola Federal de Engenharia de Itajub\'a,
 Av. BPS 1303 Pinheirinho,
 37500-903 Itajub\'a, MG -- Brazil \\
\email{lorenci@cpd.efei.br,\, newton@cpd.efei.br}
\and
 Universite Aix-Marseille III -
 Laboratoire de Mod\'elisation en M\'ecanique
 et Thermodynamique,
 U.P.R.E.S. E.A. 2596 Case 322, Avenue Escadrille
 Normandie-Niemen 13397 Marseille Cedex 20 -- France \\
\email{Henri-hugues.fliche@meca.u-3mrs.fr}
\and
 Centro Brasileiro de Pesquisas F\'{\i}sicas,
 Rua Dr.\ Xavier Sigaud 150 Urca,
 22290-180 Rio de Janeiro, RJ -- Brazil \\
\email{novello@lafex.cbpf.br}
}

\newcommand{\Fstar}{\raisebox{.2ex}{$\stackrel{*}{F}$}{}}
\renewcommand{\thefootnote}{\fnsymbol{footnote}}

\date{\today}
\abstract{
In the context of the static and spherically symmetric solution of a
charged compact object, we present the expression for the bending
of light in the region just outside the
event horizon -- the {\it dyadosphere} -- where vacuum polarization effects are
taken into account.
}
\maketitle

\section{Introduction}

The magnitude of the velocity of light described by effective nonlinear
electromagnetic theories depends on the field
dynamics. Such dependence implies an effective modification
of the flat background metric into a curved one, which is accentuated
when gravity processes are taken into account. The most famous
examples  of these aspects are the well known implications of
QED in curved spacetimes. The weak field limit of the complete one
loop QED is known as the Euler-Heisenberg Lagrangian
(Heisenberg and Euler \cite{Euler}, Schwinger \cite{Schwinger})
which yields several important results such as,  the
phenomenon of birefringence (Birula and Birula \cite{Birula},
Adler \cite{Adler}, De Lorenci et al. \cite{Lorenci}),
which describes the distinct
velocity of  light propagation for each polarization direction. There are many
works dealing with the applications of nonlinear electrodynamics,
especially, concerning
 its coupling with a gravitational field.
In this context,  Drummond and
Hathrell (\cite{Drummond})  showed the
possibility of superluminal velocities in certain spacetime configurations.
Other interesting cases can be found in
Novello et al. (\cite{Mario}), Daniels and Shore
(\cite{Shore}),
Latorre et al. (\cite{Latorre}) and Shore (\cite{Shore2}).

Recently, Ruffini (\cite{Ruffini}) and also Preparata
et al. (\cite{Preparata}) called our attention to a special region
just outside the horizon of charged black holes where the electric field goes
beyond  its classical limit, implying a situation where effects of
vacuum fluctuations should be considered. They called such region
the {\em dyadosphere}. Assuming the existence of such a region, we could
consider the possible consequences for the trajectories of light rays
that  cross it. In this work, we analyze the
consequences for the phenomenon of light bending when vacuum
fluctuation  effects are taken into account. Under such conditions, the
paths of light do not follow the usual geodesic of the gravitational
field, so it is necessary to consider the effects of the modified QED
 vacuum (Drummond and Hathrell \cite{Drummond},
Dittrich and Gies \cite{Gies}, De Lorenci et al.
\cite{Lorenci}, Novello et al. \cite{NovelloPRD} ).

In section \ref{Gravity}
we perform the coupling between  nonlinear electrodynamics and gravity.
We calculate  the correction for the Reissner-Nordstron metric from
the first contribution of the weak field limit of one loop QED.
In section \ref{IV} we present the light cone conditions for the case
of wave propagation in the Reissner-Nordstron  spacetime
modified by QED vacuum polarization effects - the dyadosphere region.
Finally, in section \ref{V}, we derive
the field equations for such a situation and evaluate its contribution
to the bending of light. Some comments on lensing effects are
presented in the appendix.

\section{The minimal coupling between gravitation and non-linear
electrodynamics}
\label{Gravity}

The Einstein gravitational field equation is given by
\begin{equation}
G_{\mu\nu} = \kappa T_{\mu\nu},
\label{Einstein}
\end{equation}
where $\kappa$ is written in terms of the Newtonian constant $G$ and
light velocity $c$ as $\kappa = 8 \pi G /c^4$.

Let us consider a class of theories defined by the general
Lagrangian $L = L(F)$, where $F = F^{\mu\nu}F_{\mu\nu}$. The corresponding
energy momentum tensor has the form
\begin{equation}
T_{\mu\nu} = -Lg_{\mu\nu} + 4L_{F}F_{\mu\alpha}F_{\nu}\mbox{}^{\alpha}
\label{g3}
\end{equation}
where $L_F$ denotes the derivative of $L$ with respect to $F$. By
considering minimal coupling of gravity to nonlinear electrodynamics,
we obtain the following equations of motion, besides Einstein equations
(\ref{Einstein}),
\begin{eqnarray}
\lefteqn{\left(L_F F^{\mu\nu}\right)_{\parallel\nu} = 0}
\label{g5}\\
\lefteqn{\Fstar^{\mu\nu}{}_{\parallel\nu} = 0}
\label{g6}
\end{eqnarray}
where double bar (${\scriptstyle \parallel}$) represents the covariant
derivative with respect  to the curved
background $g_{\mu\nu}$, and $\Fstar_{\mu\nu}$ is the
dual electromagnetic tensor. For the static and spherically symmetric
solution, the geometry is
\begin{equation}
ds^2 = A(r)dt^2 - A(r)^{-1}dr^2
-r^2d\theta^2 - r^2\sin^2\theta d\varphi^2
\label{g8}
\end{equation}
where $A(r)$ is determined by the field equations.
We set the only non zero component of the electromagnetic tensor to be
$F^{01} = f(r)$. Thus, the combined system of electromagnetism and
gravity, equations (\ref{Einstein}), (\ref{g5}) and
(\ref{g6}), reduces to the set
\begin{eqnarray}
\lefteqn{r\frac{\partial A(r)}{\partial r} + A(r)
= 1+\kappa\left[ r^2 L +4r^2 L_F f(r)^2\right]}
\label{g11}\\
\lefteqn{L_F f(r) = - \frac{Q}{4r^2}.}
\label{g13}
\end{eqnarray}
We are interested in the analysis of the weak field
limit of the complete one-loop approximation of QED, given by
the effective Lagrangian
(Heisenberg and Euler \cite{Euler}, Schwinger \cite{Schwinger})
\begin{eqnarray}
&{\cal L} &= -\frac{F}{4}
+ \frac{1}{8\pi^2}\int_{0}^{\infty} ds
\frac{{\rm e}^{-m_{\scriptscriptstyle e}^2 s}}
{s^3}
\nonumber \\
&\times&\left[\frac{e^2s^2G}{4}
\frac{{\rm Re}\cosh\sqrt{2e^2s^2(F + i G)}}
{{\rm Im}\cosh\sqrt{2e^2s^2(F + i G)}}
- \frac{2e^2s^2F}{12} -1\right]
\end{eqnarray}
where $G \doteq \Fstar_{\mu\nu}F^{\mu\nu}$.
In the limit of low frequency $\nu << m_{\scriptscriptstyle e}c^2/h$ one obtains
 the Euler-Heisenberg Lagrangian
\begin{equation}
L = -\frac{1}{4}F + \frac{\mu}{4}\left(F^2 +
\frac{7}{4} G^2\right)
\label{g15}
\end{equation}
with
\begin{equation}
\mu \doteq \frac{2\alpha^2}{45 m_{\scriptscriptstyle e}^4}.
\label{g16}
\end{equation}
Since we  only consider the electric component of $F_{\mu\nu}$
there is no contribution due to the invariant $G$.
Using Lagrangian (\ref{g15}), the integration of
equation (\ref{g11})  is
\begin{equation}
A(r) = 1 - \frac{2m}{r} - \frac{\kappa}{r}\int^r dr
\left[\frac{r^2f(r)^2}{2}
+3 \mu r^2 f(r)^4\right]
\label{g20}
\end{equation}
where, in order to set the value of the first constant of integration,
we have assumed the Schwarzschild solution in the limit of vanishing
charge. Calculating the function $f(r)$ from equation (\ref{g13}) in the
appropriate order of approximation ${\cal O}(\mu)$, one gets
\begin{equation}
f(r) = \frac{Q}{r^2} -4\mu \frac{Q^3}{r^6}.
\label{g30}
\end{equation}
Introducing this result in equation (\ref{g20})
we finally obtain the expression for $A(r)$:
\begin{equation}
A(r) = 1 - \frac{2m}{r} + \frac{\kappa Q^2}{2r^2}
- \frac{\kappa \mu Q^4}{5 r^6}.
\label{g31}
\end{equation}
The Reissner-Nordstron case arises from this solution for
the limit case $\mu = 0$.
Equations (\ref{g8}) and (\ref{g31}) yield the correct form of the
spacetime geometry  taking into account the one-loop QED in
the first order of approximation.

Since (\ref{g31}) is an approximation, it cannot be applied to the cases
for which $1/r^6 \sim 1/r^2$. Thus the contribution of the last term in
this equation is negligible for two reasons: $\mu$ is very small and
$1/r^6 << 1/r^2$.

In the next sections, however, it is shown that,
due to the non-linearity of
the electrodynamics in the dyadosphere,
there will be an additional correction
which is comparable to the Reissner-Nordstron
factor in terms of the radial
variable.

One might want to investigate the effect of
this metric on the
gravitational lensing of light propagating in the vicinity
of a charged
compact object. In the appendix we apply the formalism
developed by Frittelli et al. (\cite{Frittelli}) to the
background  spacetime defined by
equation (\ref{g8}).

\section{Light cone condition}
\label{IV}

In the case of nonlinear electrodynamics, e.g.,
Euler-Heisenberg effective theory,
the wave propagation will suffer a correction
due to vacuum polarization effects.
Such correction is usually presented in terms of a
light cone condition, which in our case, is given by
\begin{equation}
k^\alpha k^\beta g_{\alpha\beta} =
-4\frac{L_{FF}}{L_F}F^{\mu\alpha}F^{\nu}
{}_{\alpha}k_{\mu}k_{\nu}.
\label{1a}
\end{equation}
It is worth mentioning that there will be two different
modes of propagation, one for each polarization direction
(De Lorenci et al. \cite{Lorenci}). Here we will deal
only with the mode present in equation (\ref{1a}).

Condition  (\ref{1a}) can be presented in
a more appealing form as a slight
modification of the background geometry
\begin{equation}
\left(g^{\mu\nu} +  4\frac{L_{FF}}{L_F}F^{\mu\alpha}
F^{\nu}{}_{\alpha}\right)
k_{\mu}k_{\nu} = 0.
\label{1b}
\end{equation}
This property allows us to introduce the concept of an
effective geometry (Novello et al. \cite{NovelloPRD}) such that
\begin{equation}
\tilde{g}^{\mu\nu} = g^{\mu\nu} + 4\frac{L_{FF}}{L_F}F^{\mu\alpha}
F^{\nu}{}_{\alpha}
\label{1}
\end{equation}
for which $k_{\mu}$ is a null vector.
Hence, we can use the previous method to
derive the bending of light in the presence of vacuum polarization
effects due to nonlinear electrodynamics.
Using such a definition, $k_{\mu}$ is
 a null vector in the effective geometry. It can be
shown that the integral curves of the vector $k_\nu$ are geodesics.
In order to accomplish this,
an underlying Riemannian structure for the manifold
associated with the effective geometry will be required.  In other words,
this implies a set of Levi-Civita connection coefficients
$\tilde{\Gamma}^\alpha\mbox{}_{\mu\nu}=
\tilde{\Gamma}^\alpha\mbox{}_{\nu\mu}$,
by means of which there exists a covariant differential operator
(the {\em covariant derivative}), which we
denote by a semi-colon, such that
\begin{equation}
\label{Riemann}
\tilde{g}^{\mu\nu}\mbox{}_{;\,\lambda}
\equiv \tilde{g}^{\mu\nu}\mbox{}_{,\,\lambda} +
\tilde{\Gamma}^\mu\mbox{}_{\sigma\lambda}\tilde{g}^{\sigma\nu} +
\tilde{\Gamma}^\nu\mbox{}_{\sigma\lambda}\tilde{g}^{\sigma\mu}=0.
\end{equation}
From (\ref{Riemann}) it follows that
the effective connection coefficients are completely determined
from the effective geometry by the usual Christoffel formula.

Contraction of equation (\ref{Riemann}) with $k_\mu k_\nu$ results
\begin{equation}
\label{N15}
k_\mu k_\nu \tilde{g}^{\mu\nu}\mbox{}_{,\,\lambda} =
-2k_\mu k_\nu\tilde{\Gamma}^\mu\mbox{}_{\sigma\lambda}
\tilde{g}^{\sigma\nu}.
\end{equation}
Differentiating $\tilde{g}^{\mu\nu}k_\mu k_\nu = 0$, we have
\begin{equation}
\label{N16}
2k_{\mu,\,\lambda}k_\nu \tilde{g}^{\mu\nu} +
k_\mu k_\nu \tilde{g}^{\mu\nu}\mbox{}_{,\,\lambda} = 0.
\end{equation}
From these expressions we obtain
\begin{equation}
\label{N18b}
\tilde{g}^{\mu\nu}k_{\mu;\,\lambda}k_\nu = 0.
\end{equation}
Since the propagation vector $k_\mu=\Sigma_{,\,\mu}$
is a gradient, one can write
$k_{\mu;\,\lambda}=k_{\lambda;\,\mu}$.
Thus, equation (\ref{N18b}) reads
\begin{equation}
\label{geodesicb}
\tilde{g}^{\mu\nu}k_{\lambda;\,\mu}k_\nu = 0
\end{equation}
which states that $k_\mu$ is a geodesic vector.
Since it is also a null vector
in the effective geometry $\tilde{g}^{\mu\nu}$,
it follows that its integral curves are therefore
null geodesics.

\section{The influence of QED on the trajectory of light}
\label{V}

Taking the Lagrangian (\ref{g15}),
and setting the only non-zero component of electromagnetic tensor
to be $F^{01} = f(r)$, it follows that
\begin{eqnarray}
\lefteqn{F = -2 f(r)^2}
\label{9} \\
\lefteqn{L_F = -\frac{1}{4} - \mu f(r)^2}
\label{7}\\
\lefteqn{L_{FF} = \frac{\mu}{2}.}
\label{8}
\end{eqnarray}
Since we are analyzing the spherically symmetric and static solution
of a charged compact object, we set the components of the background metric to
be the same as the ones presented in equation (\ref{g8}).
Thus, the non-vanishing components of the
effective metric $\tilde{g}^{\mu\nu}$, up
to terms quadratic on the constant $\mu$, are
\begin{eqnarray}
\tilde{g}^{00} &=& A(r)^{-1}[1 + 8\mu f(r)^2]
\label{16}\\
\tilde{g}^{11} &=& -A(r)[1 + 8\mu f(r)^2]
\label{17}\\
\tilde{g}^{22} &=& g^{22}
\label{18b}\\
\tilde{g}^{33} &=& g^{33}
\label{18d}
\end{eqnarray}
where $A(r)$ is given by (\ref{g31}).
In what follows we will consider only the stronger term
arising from quantum corrections. All other terms will be
neglected. The function $f(r)$ is calculated from the equations of
the electromagnetic field, and its value is set by equation (\ref{g30}).
Using these results we obtain the line element:
\begin{eqnarray}
d\tilde{s}^2
&=& \left[1-8\mu f(r)^2\right]\left[ A(r)dt^2
- A(r)^{-1}dr^2\right] \nonumber \\
&-&r^2d\theta^2 - r^2\sin^2\theta d\varphi^2.
\label{20}
\end{eqnarray}
From the variational principle,
the following equations of motion are obtained:
\begin{eqnarray}
\lefteqn{\left[1-8\mu f(r)^2\right]A(r)\dot{t} 
= constant \doteq h_o}
\label{24}
\\
\lefteqn{r^2\dot{\varphi} = constant \doteq l_o}
\label{28}
\\
\lefteqn{\dot{r}^2 = 
\frac{h_o^2}{\left[1-8\mu f(r)^2\right]^2} -
\frac{l_o^2A(r)}{r^2\left[1-8\mu f(r)^2\right]}.}
\label{32}
\end{eqnarray}
In these equations dots mean derivatives with respect to parameter
$\tilde{s}$, and
we have adjusted the initial conditions
$\theta = \pi /2$ and $\dot{\theta} = 0$.
Performing the change of variable $r=1/v$ and expressing the
derivatives with respect to the angular variable $\varphi$ we obtain from
(\ref{24})-(\ref{32}):
\begin{equation}
{v'}^2 = \frac{h_o^2}{l_o^2}\frac{1}{[1-8\mu f(r)^2]^2} -
\frac{A(r)v^2}{1-8\mu f(r)^2}
\label{35}
\end{equation}
where $v'= dv / d\varphi$.
In the required order of approximation, functions
$A(v)$ and $f(v)$ are given by
\begin{eqnarray}
f(v) &=& Q v^2
\label{36}\\
A(v) &=& 1 - 2mv + \frac{\kappa Q^2 v^2}{2}
- \frac{\kappa \mu Q^4 v^6}{5}.
\label{37}
\end{eqnarray}
Thus, taking the derivative of equation (\ref{35}) and using the
above results, it follows that
\begin{equation}
v'' + v = 3mv^2 -\left(1 - 32 \frac{\mu h_o^2}
{\kappa l_o^2}Q \right)\kappa Q^2 v^3
+ {\cal O}(\mu^2,v^4).
\label{40}
\end{equation}
This shows that the contribution coming from QED
is of the same order of
magnitude as the classical Reissner-Nordstron charge term. One should
notice that equation (\ref{40}) could also be derived using the
formalism developed by Virbhadra et al. (\cite{Virbhadra}).

\section{Conclusion}

In this work we investigate the bending of light in the dyadosphere, a region
 just outside an event horizon of a
non-rotating charged black hole,
in which the electromagnetic field exceeds
the critical value predicted by Heisenberg and
Euler. In such region the propagation of light is affected not only by the
gravitational field, but also by the modified QED vacuum.
We show that by considering effects due to
QED, the modified metric in the first order of approximation is given by
equation (\ref{20}).

We also show that the contribution from this effect
appears in a significative order in terms
of the radial variable. Indeed, it is of
the same order of the charged term that
arises from Reissner-Nordstron
solution. Equation (\ref{40}) shows that
the correction term depends on both the
ratio $h_0/l_0$ and the charge $Q$.
Thus, for photons with the same
frequency propagating in the dyadosphere,
the effect will be stronger for
those whose trajectory is closer to
the center of attraction.

Since we obtained a contribution coming from QED of
the same order of magnitude as the classical Reissner-Nordstrom charge
term, an interesting extension of this work could be studying
gravitational lensing effects for the metric
presented in equation (\ref{20}). It is also worth analysing
the effects arising from the magnetic field due to the
rotation of a charged compact object.

\begin{acknowledgements}
This work was partially supported by {\em Conselho Nacional
de Desenvolvimento Cient\'{\i}fico e Tecnol\'ogico} (CNPq),
{\em Coordena\c{c}\~ao de Aperfei\c{c}oamento de Pessoal
de N\'{\i}vel Superior} (CAPES) and {\em Comit\'e Fran\c{c}ais
d'Evaluation de la Coop\'eration Universitaire avec le Br\'esil} (COFECUB).
\end{acknowledgements}

\appendix
\section{Comments on lensing effect}
\label{lensing}

Some interesting references about the lensing effect can be
found in Virbhadra et al. (\cite{Virbhadra}),
Kling and Newman (\cite{Kling}),
Frittelli et al. (\cite{Frittelli}) and Kling et
al. (\cite{Newman}).
Kling and Newman (\cite{Kling}), for instance,
investigated aspects of light
cones in the Schwarzschild geometry, making connections to
gravitational lensing theory and to the so-called `null surface
formulation'. Later
Frittelli et al. (\cite{Frittelli})
proposed a definition of an exact lens equation without reference
to a background spacetime and developed a formalism,
which is then applied to the Schwarzschild solution.

Following this formalism, the exact gravitational
lens equation in the framework of the metric (\ref{g8})
is given by
\begin{eqnarray}
\label{Theta}
\Theta(l,l_o,l_p)=\pm \left[\pi-2 \int_{l_o}^{l_p}\frac{dl}{\sqrt{l_p^2
 A(l_p)-l^2 A(l)}}\right.
\nonumber\\
 -\left. \int_{l}^{l_o}\frac{dl'}{\sqrt{l_p^2
 A(l_p)-{l'}^2 A(l')}}\right].
\end{eqnarray}
In the above equation, $\Theta(l,l_o,l_p)$ is the angular position of the
source with respect to the direction defined by the
observer and the lensing object (the optical axis) and $l = 1/\sqrt{2}r$ is
the inverse radial distance. It  follows that $l_o = 1/\sqrt{2}r_o$ and $l_p =
1/\sqrt{2}r_p$, where $r_o$ and $r_p$ are, respectively, the position of the
observer and the minimum distance between the light path and the lens. As
usual, the origin of the coordinate system is the position of the lens.

A generic point in the light path is given in spherical coordinates by
\begin{eqnarray}
\cos\theta&=&-\cos\theta_o\cos\Theta
+\sin\theta_o\sin\Theta\cos\gamma
\label{theta}
\\
\tan\!\varphi\!&=&\!\frac{\sin\!\varphi_o
\sin\!\theta_o \!-\!
\tan\!\Theta(\cos\!\varphi_o\sin\!\gamma \!-\!
\sin\!\varphi_o\cos\!\gamma\cos\!\theta_o)}
{\cos\!\varphi_o\sin\!\theta_o \!+\!
\tan\!\Theta(\sin\!\varphi_o\sin\!\gamma \!+\!
\cos\!\varphi_o\cos\!\gamma\cos\!\theta_o)}.
\nonumber\\
\label{phi}
\end{eqnarray}
The coordinate $\gamma$ is the azimuthal angle of the light path with respect
to the optical axis and the observer is located at $(\varphi_o,\theta_o)$.

Applying equation (\ref{Theta}) to the function $A(r)$
derived in (\ref{g31})
it follows that there is no significant
contribution from QED to the lensing
effect for the same reasons discussed
at the end of section \ref{Gravity}.


\begin{thebibliography}{}

\bibitem[1971]{Adler}
Adler, S. L. 1971, Ann. Phys. 67, 599.

\bibitem[1970]{Birula}
Bialynicka-Birula, Z \& Bialynicki-Birula, I. 1970,
{\em Phys.\ Rev.}  D2, 2341.

\bibitem[1994]{Shore}
Daniels, R. D. \& Shore, G. M. 1994,
Nuclear Phys. B425, 635.

\bibitem[2000]{Lorenci}
De Lorenci, V. A., Klippert, R.,  Novello, M \& Salim, J. M.
2000, Phys. Lett., B482, 134.

\bibitem[1998]{Gies}
Dittrich, W. \& Gies, H. 1998,
Phys. Rev., D58, 025004.

\bibitem[1980]{Drummond}
Drummond, I. T. \& Hathrell, S. J. 1980,
Phys. Rev. D22, 343.

\bibitem[2000]{Frittelli}
Frittelli, S.,  Kling, T. P. \& Newman, E. T. 2000,
Phys. Rev. D61, 064021.

\bibitem[1936]{Euler}
Heisenberg, W. \& Euler, H. 1936, {\em Z. Phys}., 98,
714.

\bibitem[1999]{Kling}
Kling, T. P. \& Newman, E. T. 1999, Phys. Rev. D59,
124002.

\bibitem[1999]{Newman}
Kling, T. P., Newman, E. T. \& Perez, A. 1999,
Phys. Rev. D61, 104007.

\bibitem[1995]{Latorre}
Latorre, J. I.,Pascual, P. \& Tarrach, R. 1995,
Nuclear Phys., B437, 60.

\bibitem[2000]{NovelloPRD}
Novello, M., De Lorenci, V. A., Salim, J. M. \& Klippert, R.
2000, Phys. Rev., D61, 045001.

\bibitem[1989]{Mario}
Novello, M. \& Jordan, S. 1989, Mod. Phys. Lett.
A4, 1809.

\bibitem[1998]{Preparata}
Preparata, G., Ruffini, R. \& Xue, S. S. 1998, A\&A, L87, 338.

\bibitem[1998]{Ruffini}
Ruffini, R. 1998, in XLIX Yamada Conference on Black Holes and
High-Energy Astrophysics, ed. H. Sato. Univ. Acad. Press.,
Tokyo.

\bibitem[1951]{Schwinger}
Schwinger, J. 1951, {\em Phys.\ Rev.} 82, 664.

\bibitem[1996]{Shore2}
Shore, G. M. 1996, Nuclear Phys., B460,
379.

\bibitem[1998]{Virbhadra}
Virbhadra, K. S., Narasimha, D. \& Chitre, S. M. 1998,
A\&A, 337, 1.

\end{thebibliography}
\end{document}